# Nanolayer and nano-convection based enhanced thermal conductivity of Copper-CO$_2$ nanofluid: A molecular dynamics approach

Zeeshan Ahmed [1], Ajinkya Sarode [1], Pratik Basarkar[1], Atul Bhargav[1] & Debjyoti Banerjee[2]
[1]Indian Institute of Technology Gandhinagar, Palaj, Gujarat, India
[2] Texas A&M University, USA
Email: zeeshan.ahmed@iitgn.ac.in

**ABSTRACT**

The use of CO$_2$ as a natural refrigerant in data center cooling, oil recovery and in CO$_2$ capture and storage which is gaining traction in recent years involves heat transfer between CO$_2$ and the base fluid. A need arises to improve the thermal conductivity of CO$_2$ to increase the process efficiency and reduce cost. One way to improve the thermal conductivity is through nanoparticle addition in the base fluid. The nanofluid in this study consists of copper (Cu) nanoparticle and CO$_2$ as a base fluid. No experimental data is available on the thermal conductivity of CO$_2$ based nanofluid. In this study, the effect of the formation of a nanolayer (or molecular layering) at the gas-solid interface on thermal conductivity is investigated using equilibrium molecular dynamics (EMD) simulations. This study also investigates the diameter effect of nanoparticle on the nanolayer, thermal conductivity and self-diffusion coefficient. In addition to this, diffusion coefficients are calculated for base fluid and nanofluid. Thickness of the dense semi-solid layer formed at the nanoparticle-gas interface is studied through radial distribution function (RDF) and density distribution around the nanoparticle. This thickness is found to increase with nanoparticle diameter. Enhancement in thermal conductivity and diffusion coefficient with nanoparticle diameter are strongly correlated, indicating that the dominant modes of heat and mass transfer are the same. The output of the current work demonstrates the enhancement in thermal conductivity due to nanoparticles addition which may improve data center cooling efficiency and CO$_2$ capture and storing.

**KEY WORDS:** Cu-CO$_2$ nanofluid, Molecular interfacial layer, Thermal conductivity, Molecular dynamic simulation

**Nomenclature**

| | |
|---|---|
| $\varphi$ | LJ potential, (J) |
| $\varepsilon$ | Interaction strength, (J) |
| $\sigma$ | Interatomic length scale between atoms, (m) |
| $r$ | Distance between two atoms, (m) |
| $E$ | Total energy of an atom, (J) |
| $F$ | Embedding energy (function of electron density) |
| $U$ | Pair potential interaction, (J) |
| $m$ | Mass of the particle |
| $\rho$ | Electron charge density, (kg/m$^3$) |
| $k_B$ | Boltzmann's constant, (J/K) |
| $N$ | Total number of atoms |
| $T$ | Thermodynamic temperature, (K) |
| $V$ | Volume, (m$^3$) |
| $J$ | Instantaneous microscopic heat flux, (W/m$^2$) |
| $f$ | Interaction between particles ruled by potential |
| $v$ | velocity of particle, (m/s) |
| $e$ | Surplus energy, (J) |
| $k$ | Thermal conductivity (W/m-K) |

**Subscripts**

| | |
|---|---|
| $i, j$ | denotes atoms |
| $\alpha, \beta$ | denotes different types of atom |

## 1. INTRODUCTION:

Global warming caused by the $CO_2$ emission from various sources has been a serious concern these days. $CO_2$ has a significant repercussion on the climate and is therefore extensively studied. Several efforts are being made to reduce carbon dioxide emission into the atmosphere by different techniques such as capture and sequestration [1, 2]. $CO_2$ capture and geological storage are considered to be feasible options to pacify and reduce greenhouse gas emissions during the transition phase towards the use of renewable energy. For this, the prognostication of thermal conductivity of $CO_2$ is of prime relevance in the process of capture, transport, and injection of $CO_2$. Further, as the drift towards environmentally tender refrigerants continues, the performance and cost analysis of $CO_2$ as a natural refrigerant has drawn the attention of many researchers across the globe. According to Y. Solemdal et al. [3], $CO_2$ systems result in reduced energy consumption in the referred system, which in turn results in improved system performance and reduced annual cost.

Nanofluids are defined as fluids with suspensions of nanoparticles. In comparison to the base fluids, these are potential heat transfer fluids which, even at very low concentrations, have exhibited an exceptional increase in thermal conductivities. For example, thermal conductivity of

Al$_2$O$_3$-water nanofluid was increased by 10% [4] and 30% [5] with diameters of 13 nm and 40 nm respectively at the same particle volume fraction of 4.3%, and up to 40% increase if the Cu particles of 10 nm at a much lower concentration of 0.3% were dispersed in water [6]. Even more interesting is the finding that there is up to 150% thermal conductivity enhancement in a suspension of 1.0 vol. % multiwalled CNTs in oil [7]. Till date, numerous studies have been done on determining the thermal conductivity of nanofluids. Furthermore, Jang and Choi [8] proposed four mechanisms of energy transport in nanofluids: (a) base fluid molecules interactions, (b) thermal diffusion of nanoparticles in base fluid, (c) Brownian motion of nanoparticles and (d) thermal interactions of nanoparticles with base fluid molecules. Ren et al. [9] proposed a collateral model describing the heat transfer mechanism in nanofluid by adding a nanolayer effect on thermal conductivity. The motion of the nanoparticle enhances the overall heat transport by micro-convection in the suspended fluid. Prasher et al. [10] captured the aggregation effect in addition to Brownian induced convection for the enhancement. Such behavior and physical phenomenon makes nanofluids a potential candidate to enhance the heat transfer properties. Lee et al. [11] studied the effect of particle size on Cu-liquid argon based nanofluid with different volume fraction and found out that there is reduction in thermal conductivity enhancement with decrease in nanoparticle diameter. Artificial correlations which arise from single nanoparticle systems and periodic boundary conditions result in very high values of thermal conductivities as suggested by MG Muraleedharan et al. [12]. This can be mitigated by considering a multi-nanoparticle system with smaller diameter or single-nanoparticle system with larger nanoparticle

Extensive research manifests the adverse effect of nanoparticle exposure on human health [13]. But in this study, the use of gas based nanofluids is in space application, in nuclear power plants and as a refrigerant in data center cooling where the system is closed. So, these nanoparticles suspended in gas are not exposed to the environment and hence do not affect human health.

To the best of the author's knowledge, thermal conductivity of nanofluid in gas phase is not reported in open literature so far. Several studies have been done broadly on the diffusion of nanoparticles and transport of nanoparticles in gases which proves the presence of nanoparticle in the gaseous phase [14-17]. In this work, molecular dynamics (MD) simulation is performed using Large-scale Atomic/Molecular Massively Parallel Simulator (LAMMPS) [18] to study the effect of nanolayer and Brownian motion on the enhancement of thermal conductivity of novel Cu-CO$_2$ nanofluids. To study the diameter effect on the thermal conductivity of the nanofluid, nanoparticles having 1 nm, 2 nm and 3nm diameters are considered, keeping the volume fraction constant at 1.413%. This study predicts the presence of nanolayer in the gaseous phase and can be considered as one of the reasons for enhanced thermal conductivity, while the other reason being the augmented self-diffusion coefficient of gas molecules in the nanofluid.

## 2. SIMULATION SETUP AND METHODOLOGY:

Molecular dynamics (MD) simulation is being increasingly adopted as a tool to perform preliminary assessments of nanoparticle (NP) fluid interactions and determining thermal, mechanical and other properties of interest. The potential interaction between the atoms is calculated through potential energy function which further estimates the force acting on them. This potential energy function depends on the position of individual atoms present in the simulation domain and is composed of bonded and non-bonded energy interactions. The bonded interaction includes energy stored due to the bond-stretching, angle of bending. The non-bonded interactions are evaluated from the Van der Waals and the electrostatic interactions (eg. Coulombic) are calculated using the Particle Mesh Ewald method [19].

In the present work, gaseous $CO_2$ with a suspension of Cu nanoparticle is modeled using LAMMPS. The $CO_2$ molecules are represented by the conventional EPM2 model [20] as it predicts thermodynamic properties better. The LJ potential [20] is used for interatomic interaction between different atoms, which is given by:

$$\varphi(r_{ij}) = 4\varepsilon_{ij}\left[\left(\frac{\sigma_{ij}}{r_{ij}}\right)^{12} - \left(\frac{\sigma_{ij}}{r_{ij}}\right)^{6}\right] \quad (1)$$

The cutoff radius of ~$4\sigma_{O-O}$ is chosen because thermal conductivity is almost independent after this distance. For EPM2 $CO_2$, the LJ potential parameters are shown in Table 1.

Table 1. LJ parameters used for carbon-carbon and oxygen oxygen interaction

| $\sigma_{O-O}$ | $\varepsilon_{O-O}$ | $\sigma_{C-C}$ | $\varepsilon_{C-C}$ |
|---|---|---|---|
| 3.03 Å | 80.507 K | 2.757 Å | 28.13 K |

The Lorentz Berthelot mixing rule [18] is used to compute the interaction between different types of atoms $i$ and $j$, which is given by:

$$\sigma_{ij} = \frac{\sigma_i + \sigma_j}{2} \quad (2)$$

$$\varepsilon_{ij} = \sqrt{\varepsilon_i \times \varepsilon_j} \quad (3)$$

For interactions between Cu atoms, EAM potential is used [18]. In EAM potential, the potential energy of an atom, $i$, is given by:

$$E_i = F_\alpha\left(\sum_{i \neq j} \rho_\beta(r_{ij})\right) + \frac{1}{2}\sum_{i \neq j} U_{\alpha\beta}(r_{ij}) \quad (4)$$

Molecular dynamics simulations are performed in the canonical ensemble (NVT) and visualized by Visual Molecular Dynamic (VMD) [21]. To validate the simulation method with experimental data, we calculated the thermal conductivity of base fluid, i.e. gaseous $CO_2$ at $T=300K$ and $\rho = 186 kg/m^3$ through Green- Kubo formalism, which gave a validation error of 2.8%. The Nose–Hoover thermostat was used for maintaining the constant temperature conditions of the whole system. Spherical region is carved out by inserting Cu nanoparticle in three different configurations. Figure 1 shows the simulation box of 2 nm nanoparticle with 755 molecules and the other configurations details are shown in the Table 2.

Table 2. Details of other configurations used to perform the simulations.

| NP diameter (nm) | Box dimension (Å) | Number of molecules | Volume Fraction (%) |
|---|---|---|---|
| 1 | 33.33 | 100 | 1.413 |
| 2 | 66.66 | 755 | 1.413 |
| 3 | 100 | 2510 | 1.413 |

The size of the simulation domain having periodic boundary conditions in all the three directions is varied from 33.33 Å to 100 Å to have constant bulk density of gaseous $CO_2$ (186 kg/m$^3$) and constant volume fraction of 1.413% for all configurations. The two phases (i.e. gas and solid) present in the domain are grouped separately. Minimization is done to remove close contacts and thus avoid high potential energy collisions. Sufficient time steps were performed for the equilibration process to achieve equilibrium state separately for each of the individual grouped systems while the other was kept immobile, under the micro-canonical ensemble (NVE) and Langevin thermostat. The canonical ensemble is used for the whole system before switching to NPT. The pressure and temperature is fixed at 1atm and 300 K, respectively. Then, fluctuation of autocorrelations is performed under the micro-canonical ensemble (NVE) for data computation to calculate thermal conductivity for each nanofluid system. The same procedure was followed for all the three systems. Newton's equations of motion were integrated using the velocity Verlet algorithm [22] with a sufficient time step.

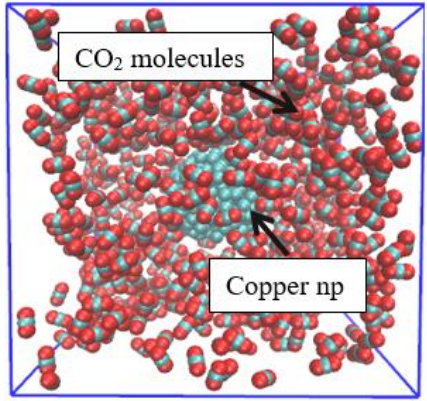

Figure 1: Cross sectional view of the Cu-CO$_2$ nanofluid with 2 nm diameter under investigation.

The mean square displacement (MSD) is calculated for gas molecules and solid nanoparticles using the atoms position at different intervals of time. The distance moved by an atom/molecule is measured using MSD, which is defined as:

$$MSD(t) = \left\langle \Delta \vec{r_i}(t)^2 \right\rangle = \left\langle (\Delta \vec{r_i}(t) - \Delta \vec{r_i}(0))^2 \right\rangle \quad (5)$$

where, $\vec{r_i}(t)$ is the position of $i^{th}$ atom at time $t$ and ($\vec{r_i}(t) - \vec{r_i}(0)$) is the displacement of $i^{th}$ atom, over a time interval t. An estimate of the number of colliding atoms in the simulation is determined by MSD.

MD method relates the thermal conductivity of fluid to equilibrium heat flow autocorrelation function through Green-Kubo equation [23], which is written as:

$$k = \frac{V}{3k_B T^2} \int_0^\infty \langle J(0).J(t) \rangle dt \quad (6)$$

and $J$ is the instantaneous microscopic heat flux vector given by:

$$J = \frac{1}{V}\left[\left[\sum_{j=1}^{N} e_j v_j - \sum_{\alpha=1}^{2} h_\alpha \sum_{j=1}^{N_\alpha} v_{\alpha j}\right] + \frac{1}{2}\left[\sum_{i=1}^{N}\sum_{j=1, j\neq i}^{N} r_{ij}(v_j.F_{ij})\right]\right] \quad (7)$$

and $e_j$ represents surplus energy of the atom $j$, which is calculated by:

$$e_j = \sum_j \frac{1}{2} m_j v_j^2 + \frac{1}{2}\sum_{i \neq j} \varphi_{ij} \quad (8)$$

where $v_j$ is the $j^{th}$ particle velocity, $h_\alpha$ is the average partial enthalpy of species $\alpha$, $F_{ij}$ and $r_{ij}$ are the interatomic forces and distance between $i^{th}$ and $j^{th}$ particles, respectively, $N_\alpha$ is the number of particles of kind $\alpha$ and $N$ is the total number of particles. Average partial enthalpy is the sum of average kinetic energy, potential energy and interaction potential term, which is given by:

$$h_\alpha = \frac{1}{N_\alpha}\sum_{j=1}^{N_\alpha}(e_j + r_j . F_j) \qquad (9)$$

To calculate thermal conductivity of multicomponent system, $h_\alpha$ is an important factor to consider [24]. In a pure fluid, $h_\alpha$ is always zero for a single-component system due to the zero-average velocity, but it is non-zero for multi-component systems. The total energy flux is the sum of energy transfer due to mass flow, boundary (pressure) work, and heat conduction. Since the objective is to calculate thermal conductivity, only conduction energy flux should be considered. Hence, the term containing $h_\alpha$ should be subtracted from equation (6) to avoid anomalous high thermal conductivity in multi-component systems.

## 3. RESULTS AND DISCUSSION:

### 3.1 Density Distribution:

The nanolayer surrounding the nanoparticle shown in Figure 2 can be considered as a region having higher density than the bulk fluid density. The thickness of the nanolayer can be estimated by analyzing the density distribution of $CO_2$ molecules within the domain. For this, the computational domain is divided into several spherical bins and average density of $CO_2$ molecules in each bin is plotted as shown in Figure 3.

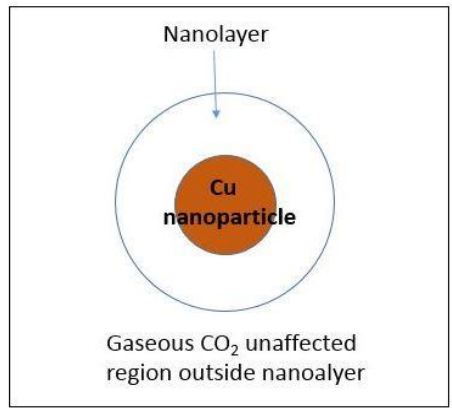

Fig.2 Schematic of the Cu nanoparticle and nanolayer formation around the nanoparticle

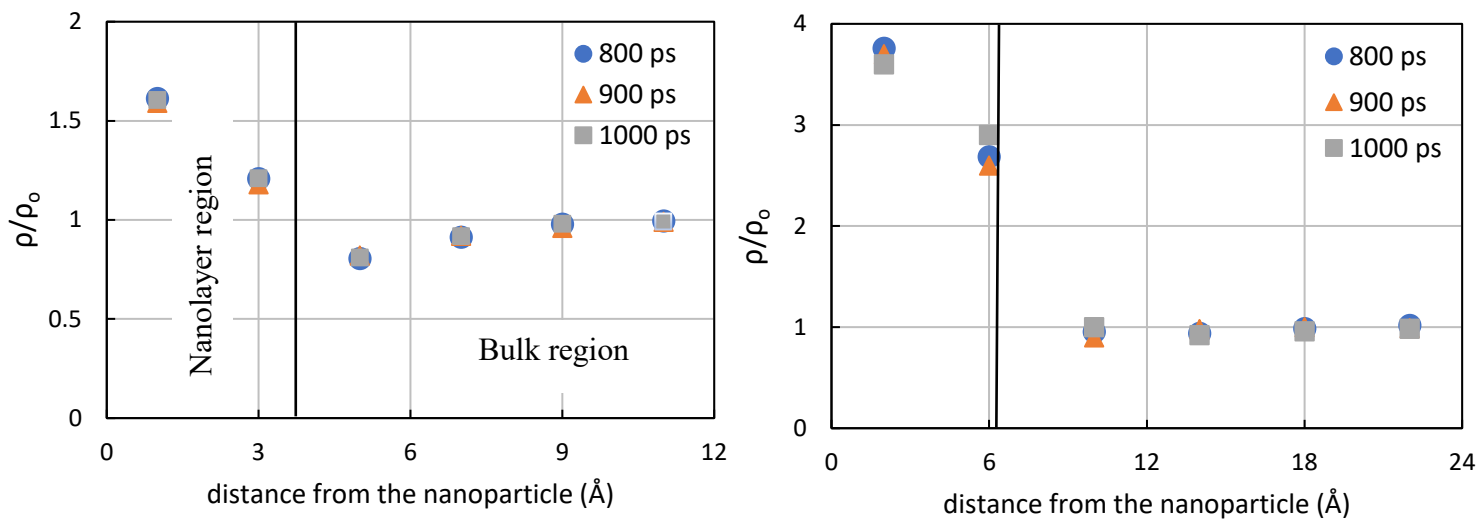

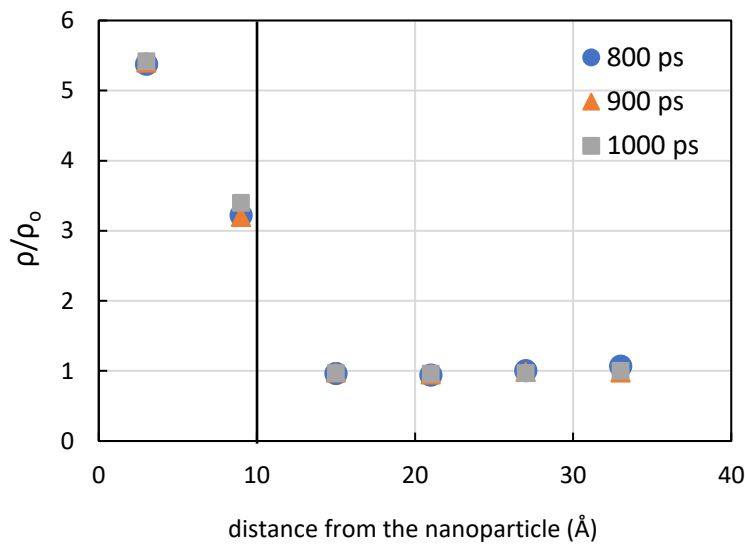

(c)

Figure 3: (a), (b) and (c) show the density distribution of $CO_2$ for 1nm, 2nm and 3nm nanoparticle diameter.

Figure 3 shows that near the nanoparticle, the density of $CO_2$ is highest when compared to the bulk density. Based on the density distribution of $CO_2$ molecules, highest thickness of nanolayer is observed for 3 nm diameter nanoparticle and lowest for 1 nm diameter nanoparticle. Therefore, thickness of the nanolayer can be considered to be a function of nanoparticle diameter. Table 3 shows the semi-solid nanolayer thickness for different nanoparticle diameter. Also, the thickness of the nanolayer remains approximately the same with time for all nanoparticles. This justifies the movement of the nanolayer along with Brownian motion of the nanoparticle. The same is visualized in VMD and a similar observation was also reported by Li et al. [25].

Table 3. Nanolayer thickness values for different nanoparticle diameter

| NP diameter (nm) | Nanolayer thickness (nm) |
|---|---|
| 1 | 0.4 |
| 2 | 0.7 |
| 3 | 1 |

## 3.2 Radial distribution function:

The radial distribution function (RDF) for a solid–gas interface with different nanoparticle diameters is shown in Figure 4. The addition of nanoparticle changed the structure of the gas near the interface to a denser form as compared to the RDF of bulk gas. From Figure 4, the nanolayer surrounding 3 nm particle is denser than that of 2 nm and 1 nm particles. Among the three configurations studied, the number of $CO_2$ molecules interacting with the nanoparticle is higher for larger particle diameter. This supports the variation in the peak density observed for different diameter nanoparticles as shown in Figure 3.

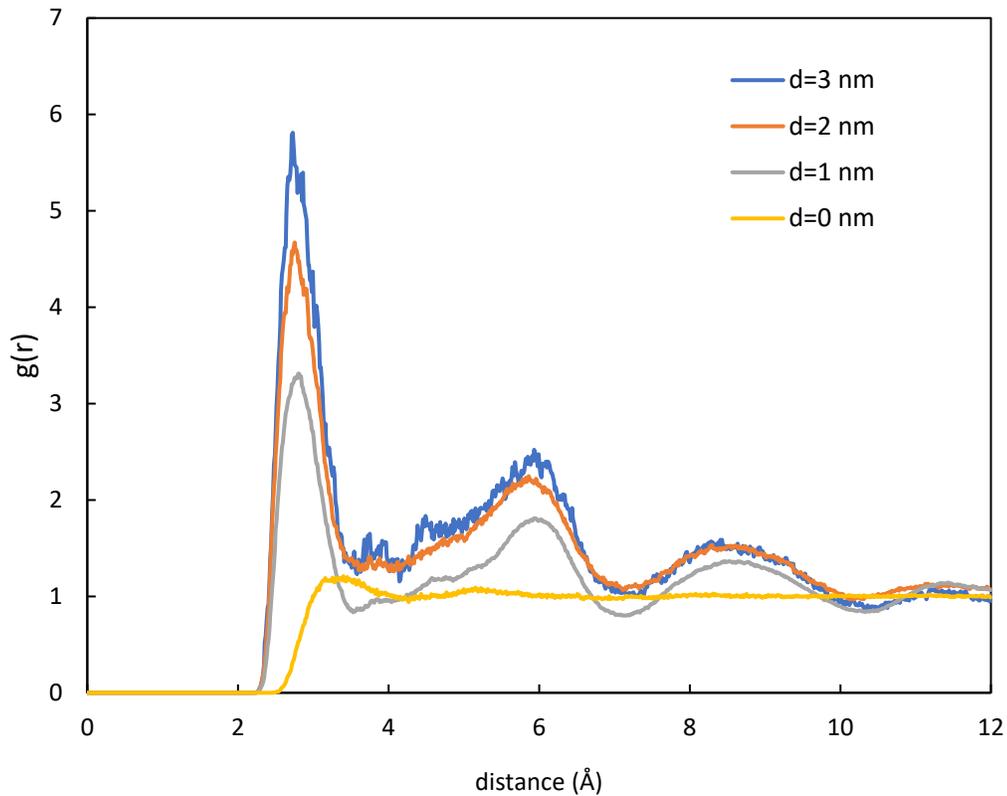

Figure 4: RDF for bulk gas (d=0) and nanofluids with particle diameters of 1nm, 2nm and 3nm as a function of radial distance from nanoparticle.

## 3.3 Mean square displacement (MSD):

The MSD of $CO_2$ gas molecules in nanofluid and the MSD of nanoparticles are calculated and compared with the MSD of the base fluid (i.e. $CO_2$ gas) as shown in Figure 5. As depicted from Figure 5, Brownian motion of the nanoparticles is less and hence, it can be assumed that they are slow in transporting heat. It is observed that the MSD of gas molecules in nanofluid is higher compared to the base fluid. The non-bonded interactions between $CO_2$ molecules present in the nanolayer and the bulk fluid increases with nanoparticle diameter due to more $CO_2$ molecules in the nanolayer. Hence, it is observed that there is an increase in MSD for a nanofluid having larger particle diameter. This increased movement of the gas molecules creates localized nano-convection.

The Einstein's relation [26] is used to calculate the self-diffusion coefficient (D) which is given as:

$$\lim_{t \to \infty} \frac{\left\langle \Delta \vec{r_i}(t)^2 \right\rangle}{3t} = 2D \qquad (10)$$

Using equation 10, the self-diffusion coefficient of the base fluid, and gas molecules in different nanofluid is calculated and compared with the diffusion coefficient of pure $CO_2$ gas as shown in Figure 7. It is observed that for nanofluids, the self-diffusion coefficient increases almost linearly with increase in particle diameter. These findings are in good agreement with the Ar-Cu nanofluid results shown by Sarkar and Selvam [26].

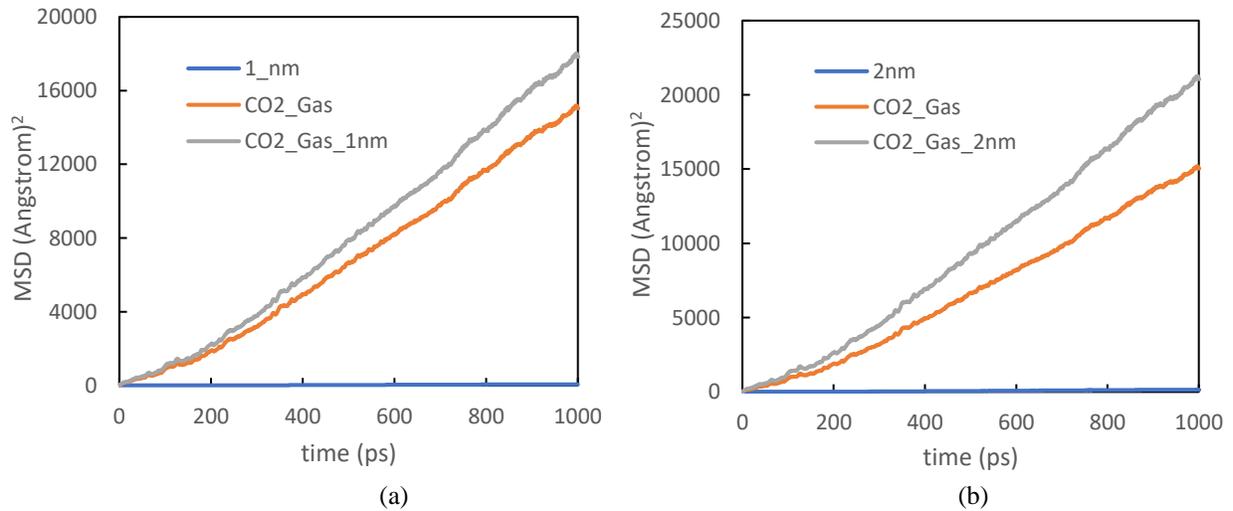

(a)      (b)

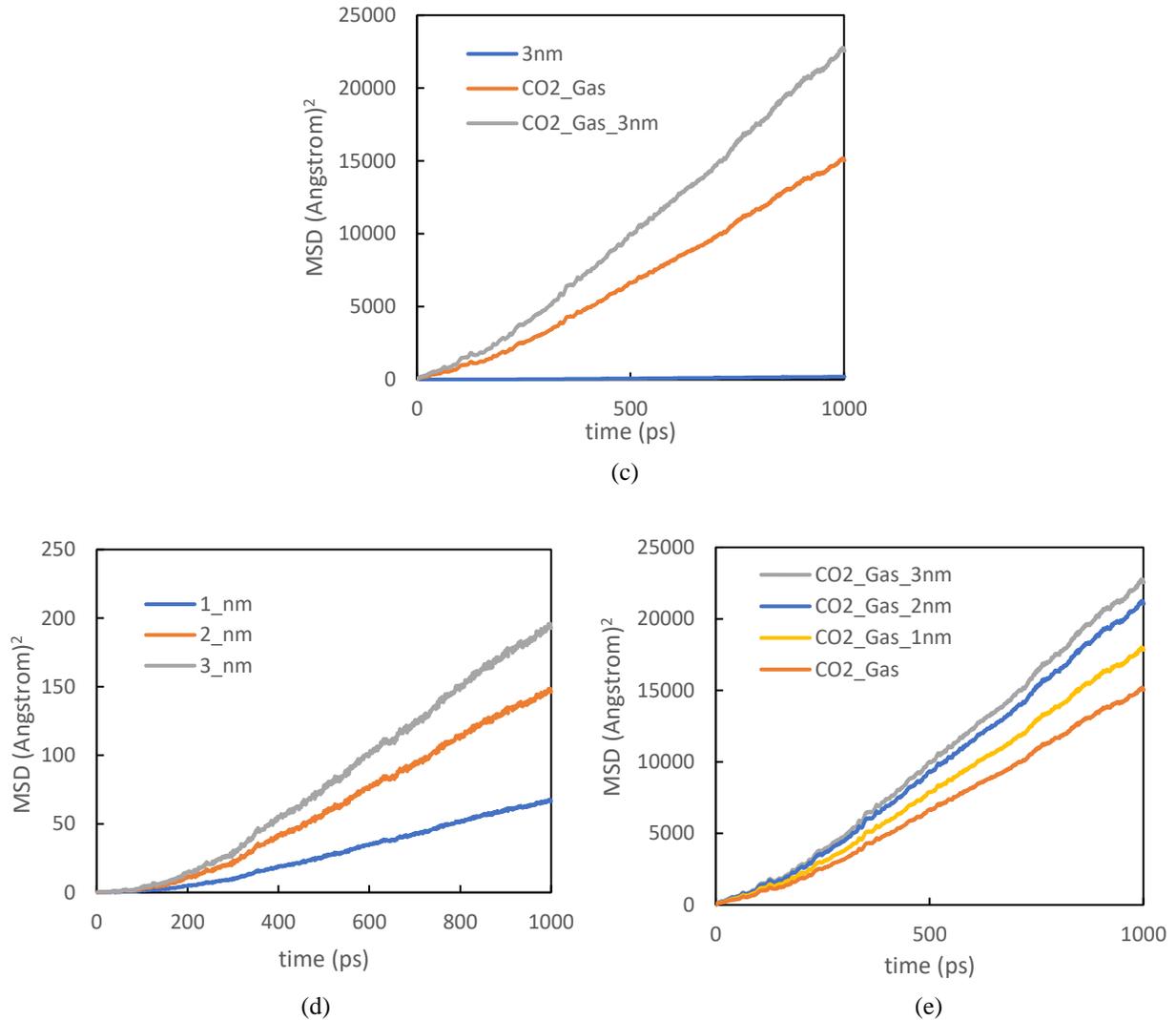

Figure 5: (a), (b) and (c) compares the MSD of 1nm, 2nm and 3nm solid particles, MSD of base fluid and MSD of nanofluid containing 1nm, 2nm and 3nm particles. (d) compares the MSD of 1nm, 2nm and 3nm diameter. (e) compares the MSD of the base fluid and the nanofluid with different diameters.

The thermal conductivity of nanofluid is calculated via Green-Kubo theory with different nanoparticle loadings by keeping a constant volume fraction of 1.413%. Accuracy of the computed thermal conductivities is determined by observing the convergence of thermal conductivities with time. Figure 6 shows the converged thermal conductivity value for different nanofluid systems. From Figure 7, it is obvious that the thermal conductivity of liquid in nanofluid is higher than that of base fluid and it has an increasing trend with the increase in nanoparticle diameter. The solid's contribution (Cu in this study) to nanofluid's thermal conductivity is negligible, no matter how high the thermal conductivity of solid nanoparticle [23]. It is the lattice thermal conductivity, not the electronic thermal conductivity, which enhances the nanofluid thermal conductivity as there is vibrational transportation of heat between nanoparticle and base fluid. The thicker and denser the nanolayer, enhancement in the thermal conductivity is more significant. The increase in thermal conductivity is due to the increase in localized nano-convection and presence of nanolayer in our current nanofluid system. Figure 7 shows a linear correlation between the thermal conductivity and diffusion coefficient enhancement.

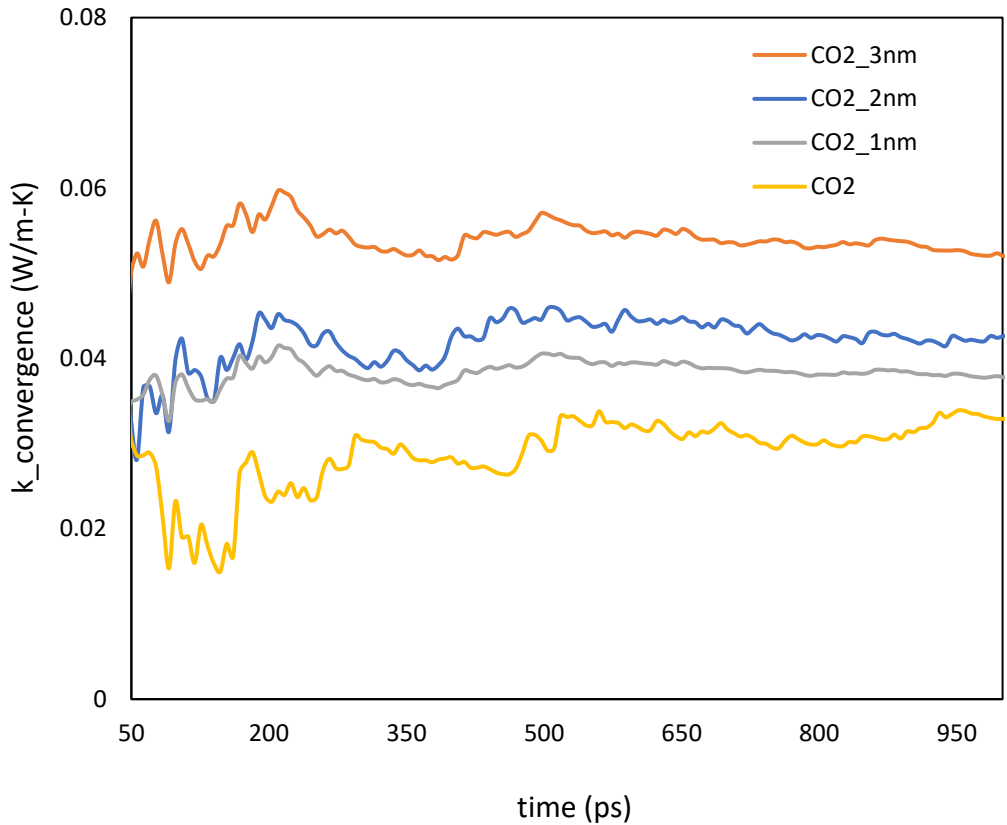

Figure 6: Convergence plot of thermal conductivity for base fluid and different nanoparticle diameter nanofluid

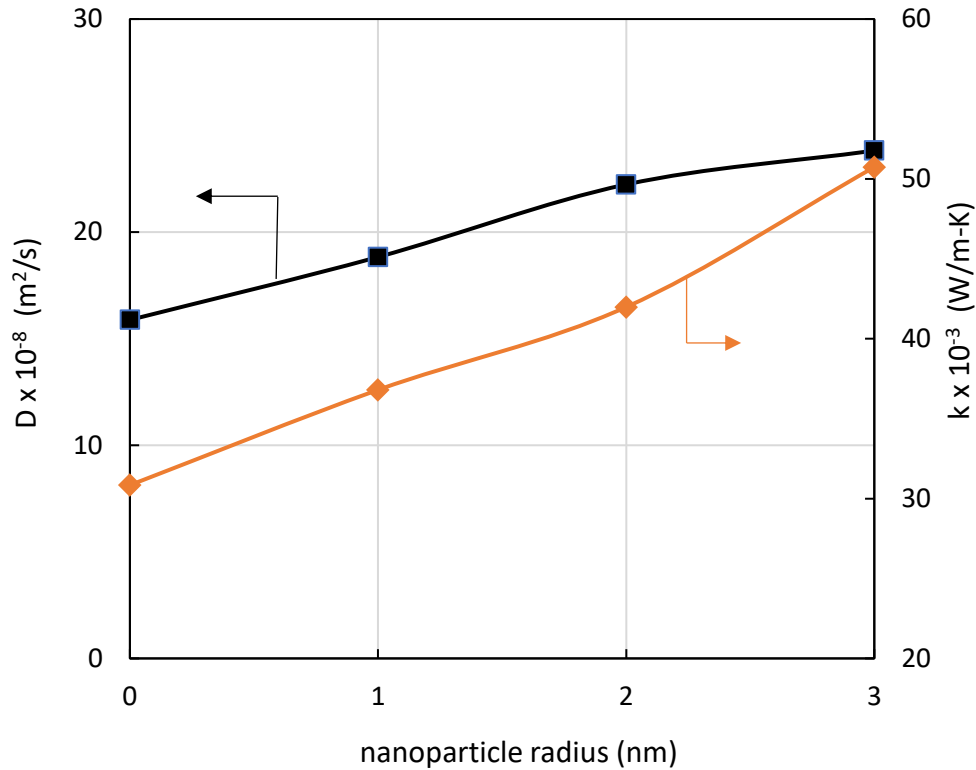

Figure 7: Self- Diffusion coefficient and thermal conductivity variation with nanoparticle diameter

**CONCLUSION:**

MD simulations were performed to study the diameter effect on the thermal conductivity with constant volume fraction of 1.413% using Green Kubo formalism. A dense layer of $CO_2$ molecules around the nanoparticle is visualized by VMD and its thickness in three different cases is calculated. Density distribution around the nanoparticle is done to determine the nanolayer thickness. The nanolayer thickness for 1 nm, 2 nm, and 3 nm particle diameter systems are 0.4 nm, 0.7 nm, and 1 nm respectively. It is also seen that the effect of the nanoparticle is very low at a distance further away from it. The density of the system is same as bulk density of gaseous $CO_2$ (186 kg/m$^3$). The radial distribution function (RDF) for a solid–gas interface with different nanoparticle diameters is compared with the RDF of bulk gas and it is shown that the bulk structure of the gas changed with the addition of nanoparticle and it apparently became denser near the nanoparticle. MSD analysis shows that the presence of Cu particle leads to an increased motion of the gas molecules and hence, enhanced the self-diffusion coefficient of gas molecules in the nanofluid. The thermal conductivity enhancement of the nanofluid is 22%, 39%, and 68% for 1 nm, 2 nm, and 3nm diameter, respectively. The enhancement can be attributed to the localized nano-convection and nanolayer formation. Hence, gas-based nanofluids are like liquid-based nanofluids, which have the potential to increase the thermal conductivity of the base gas and therefore, enhance the heat transfer performance of the gas. Improved heat transfer performance of gas coolants can have a significant impact on the performance and safety of gas-cooled nuclear power reactors, in space applications, as a refrigerant in data center cooling, as well as, in many other technologies.